\documentstyle[amssymb,preprint,aps]{revtex}

\begin{document}
\title{Collision of Electromagnetic Shock Waves Coupled with Axion Waves: An Example}
\author{M.Halilsoy\thanks{%
e-mail: mustafa.halilsoy@emu.edu.tr} and I.Sakalli\thanks{%
e-mail: izzet.sakalli@emu.edu.tr}}
\address{Physics Department, Eastern Mediterranean University, (EMU), G.Magosa,\\
Mersin 10 (N.Cyprus), Turkey. }
\date{\today}
\maketitle

\begin{abstract}
We present an exact solution that describes collision of electromagnetic
shock waves coupled with axion plane waves. The axion has a rather special
coupling to the cross polarization term of the metric. The initial data on
the null surfaces is well-defined and collision results in a singularity
free interaction region. Our solution is a generalization of the
Bell-Szekeres solution in the presence of an axion field.
\end{abstract}

\section{Introduction}

In general relativity colliding waves that yield non singular spacetime to
the future of collision are very few so far. It is known that in most cases
the collision creates an all-encompassing spacelike singularity. Examples to
the rare former class are the colliding gravitational wave solution of
Chandrasekhar and Xanthopoulos (CX) [1], colliding electromagnetic (em)
shock waves solution of Bell and Szekeres (BS) [2] and its cross-polarized
extension [3,4]. These solutions share the common feature of admitting a
Cauchy horizon (CH) instead of a singular surface. Later on detailed
perturbation analysis of the BS spacetime revealed that the CH formed turns
out to be unstable [5,6]. This property has been verified in exact solutions
by incorporating fields such as gravity, dilaton and scalar fields [7-9].
Particular finding was that certain types of scalar fields in colliding
waves make things worse, i.e. they convert CH into scalar curvature
singularities. Compared with the milder types of singularities such as
non-scalar curvature \ and quasiregular types, this type of singularity is
the strongest. We maintain, however, that some scalar fields preserve CHs
without converting them into spacetime singularities. Singularity analysis
of the BS solution has been considered by Matzner and Tipler [10], Clarke
and Hayward[11] and more recently by Helliwell and Konkowski\ [12] (and also
references cited there in). Clarke and Hayward in particular gave a detailed
exposition of the global structure of the BS spacetime by classifying the
singularity as quasiregular, i.e. of topological character, which is the
mildest type among singularities.

In this paper we present a new colliding wave solution in the
Einstein-Maxwell-Axion-Dilaton theory which is valid only in the limit of
vanishing dilaton with a special coupling constant and amplitude for the
axion. Our solution has similar properties to the BS solution, albeit in
addition to the em field we have an axion and gravitational waves
``impulse+shock'' created by \ the presence of the axion. The nice feature
of our solution is that we have a well-posed, physical Cauchy data on
intersecting null surfaces. As a result the interaction region emerges, in
analogy with the BS solution free of singularities. Our metric is a cross
polarized one and the axion is coupled within the cross polarization term.
Linear polarization limit of the metric removes the axion and brings us back
to the BS\ solution. Whether this particular observation has astrophysical
relevance or not remains to be seen [13].Finally, we apply a coordinate
transformation to our metric to reveal its anti de Sitter (AdS) structure.
AdS property is encountered in the throat limit of an extreme
Reissner-Nordstorm black hole which is given by the direct product $%
AdS_{2}\times S^{2}$ known as the Bertotti-Robinson (BR) spacetime.

Organisation of the paper is as follows: Section II presents the
Einstein-Maxwell-Axion-Dilaton problem and its solution with all physical
and geometrical quantities given in the Appendices A and B. Section III
formulates the problem as a collision and gives the transformation to the BR
form. We conclude the paper with section IV.

\bigskip

\section{Solution for Axion Coupled Em Fields}

We start with a general action which involves a dilaton field as well

\begin{equation}
S=\frac{1}{16\pi }\int \left| g\right| ^{\frac{1}{2}}d^{4}x\left\{
-R+2\left( \partial _{\mu }\phi \right) ^{2}+\frac{1}{2}e^{4\phi }\left(
\partial _{\mu }\kappa \right) ^{2}-e^{-2\phi }F_{\mu \upsilon }F^{\mu
\upsilon }-\kappa F_{\mu \upsilon }\widetilde{F}^{\mu \upsilon }\right\} 
\eqnum{1}
\end{equation}

where $\phi $\ is the dilaton, $\kappa $\ is the axion and $F_{\mu \upsilon
} $\ is the em field tensor.\ The duality operation is defined by $%
\widetilde{F}^{\mu \upsilon }=\frac{1}{2}\left| g\right| ^{-\frac{1}{2}%
}\epsilon ^{\mu \upsilon \alpha \beta }F_{\alpha \beta }$ where $\epsilon
^{0123}=1$ with $x^{\mu }=\left( u,v,x,y\right) .$ The em field $2-$form $F$
is generated from the potential $1-$form by $F=dA,$ where $A_{\mu }$ is an
Abelian vector field. Although our starting point is rather general, in the
sequel we shall set the dilaton field to zero (or a constant) and only in
such a limit our solution will be valid. Our line element is represented in
most compact form by

\begin{equation}
ds^{2}=2dudv-\Delta dy^{2}-\delta \left( dx+q_{0}\tau dy\right) ^{2} 
\eqnum{2}
\end{equation}

in which our notation is as follows

\[
\Delta =1-\tau ^{2} 
\]

\[
\delta =1-\sigma ^{2} 
\]

\begin{equation}
\tau =\sin \left( au\Theta \left( u\right) +bv\Theta \left( v\right) \right)
\eqnum{3}
\end{equation}

\[
\sigma =\sin \left( au\Theta \left( u\right) -bv\Theta \left( v\right)
\right) 
\]

\[
q_{0}=\text{ }const. 
\]

The $\left( u,v\right) $ are obviously the null coordinates, $\left(
a,b\right) $ are the constant em parameters and $\left[ \Theta \left(
u\right) ,\Theta \left( v\right) \right] $ stand for the step functions.

In the Appendix A, we give our choice of null tetrad and all the physically
relevant quantities. Let us note that we have inserted the step functions
for the later convenience to prepare the ground for the problem as a problem
of colliding waves. Suppressing the step functions naturally removes the
Dirac delta function terms, $\delta \left( u\right) $ and $\delta \left(
v\right) $, in Appendix A. In particular this spacetime (for $u>0,v>0$, i.e.
excluding the boundaries so that impulsive delta function terms are omitted)
satisfies $9\Psi _{2}^{2}=\Psi _{0}\Psi _{4}$ which implies that it is a
special type-D spacetime. For $q_{0}=0$ , it reduces to the BS spacetime
representing colliding pure em waves. We wish now to have $q_{0}\neq 0$ and
consider the field equations of the above action. It will turn out, however,
that only for the specific parameter $q_{0}=1$ the axionic contribution will
match the deficiency account of the energy-momentum and we shall have an
acceptable solution. Variational principle yields the field equations

\[
\nabla _{\mu }\left( F^{\mu \upsilon }+\kappa \widetilde{F}^{\mu \upsilon
}\right) =0 
\]

\begin{equation}
\Box \kappa =-F_{\mu \upsilon }\widetilde{F}^{\mu \upsilon }  \eqnum{4}
\end{equation}

\[
2\Box \phi =F_{\mu \upsilon }F^{\mu \upsilon }+\left( \nabla \kappa \right)
^{2} 
\]

\[
\left( \Box \text{: the covariant Laplacian }\right) 
\]

\bigskip together with the Einstein equations $\left( c=1=G\right) $

\begin{equation}
G_{\mu \upsilon }=-8\pi T_{\mu \upsilon }^{total}  \eqnum{5}
\end{equation}

We find that the following choice, together with line element (2),
constitutes a solution to the problem

\[
A_{\mu }=-\frac{1}{\sqrt{2}}\sin \left( au-bv\right) \left[ \delta _{\mu
}^{x}+\delta _{\mu }^{y}\sin (au+bv)\right] 
\]

\begin{equation}
\kappa =\sin (au-bv)  \eqnum{6}
\end{equation}

\[
\phi =0\text{ \ \ \ } 
\]

\[
q_{0}=1 
\]

where we have suppressed the step functions. Invariants of the vector field $%
A_{\mu }$ are

\begin{eqnarray}
F_{\mu \upsilon }F^{\mu \upsilon } &=&2ab\cos ^{2}\left( au-bv\right) 
\eqnum{7} \\
F_{\mu \upsilon }\widetilde{F}^{\mu \upsilon } &=&-4ab\sin (au-bv)  \nonumber
\end{eqnarray}

which vanish for both $a=0$ and $b=0$ . We recall that in the pure em
problem (without axion, or $q_{0}=0$ ) we have

\begin{eqnarray}
F_{\mu \upsilon }F^{\mu \upsilon } &=&2ab  \eqnum{8} \\
F_{\mu \upsilon }\widetilde{F}^{\mu \upsilon } &=&0  \nonumber
\end{eqnarray}

i.e. the invariants are both constant while here they are variables. The
energy-momentum tensor of the axion that we adopt here is

\begin{equation}
4\pi T_{\mu \upsilon }^{A}=\frac{1}{12}\left( 3H_{\mu \lambda \kappa
}H_{\upsilon }^{\text{ \ \ }\lambda \kappa }-\frac{1}{2}g_{\mu \upsilon
}H_{\alpha \beta \lambda }H^{\alpha \beta \lambda }\right)  \eqnum{9}
\end{equation}

The anti-symmetric tensor $H^{\mu \upsilon \lambda }$ is expressed in terms
of the scalar field $\kappa $ by

\begin{equation}
H^{\mu \upsilon \lambda }=\epsilon ^{\alpha \mu \upsilon \lambda }\kappa
_{,\alpha }  \eqnum{10}
\end{equation}

so that the axion energy-momentum tensor is expressed by

\begin{equation}
4\pi T_{\mu \upsilon }^{A}=\frac{1}{2}\left( \kappa _{,\mu }\kappa
_{,\upsilon }-\frac{1}{2}g_{\mu \upsilon }\left( \nabla \kappa \right)
^{2}\right)  \eqnum{11}
\end{equation}

The equality of $-8\pi T_{\mu \upsilon }^{total\text{ \ }}$(or the $G_{\mu
\upsilon }$) to the sum of the energy-momenta due to the em and axion fields
is satisfied by the expressions given in Appendix B, verifying the solution
above.

\section{Colliding Wave Formulation of the Problem}

Since the null coordinates are \ already introduced with the step functions
the formulation of the problem as a collision follows in a simple manner.
From the right (the $u$-dependent, Region II), the line element, incoming em
field $A_{\mu }(u)$ and the axion $\kappa (u)$ are given respectively by

\[
ds^{2}=2dudv-\cos ^{2}\left( au\Theta (u)\right) \left[ dy^{2}+\left(
dx+\sin \left( au\Theta (u)\right) dy\right) ^{2}\right] 
\]

\[
A_{\mu }(u)=-\frac{1}{\sqrt{2}}\sin \left( au\Theta \left( u\right) \right) %
\left[ \delta _{\mu }^{x}+\delta _{\mu }^{y}\sin \left( au\Theta (u)\right) %
\right] 
\]

\begin{equation}
\kappa (u)=\sin \left( au\Theta (u)\right)  \eqnum{12}
\end{equation}

The Cauchy data in this region is accompanied by the gravitational wave

\begin{equation}
\Psi _{4}\left( u\right) =\frac{a}{2}\left[ -i\delta \left( u\right)
+a\Theta (u)\left( \cos ^{2}\left( au\right) +3i\sin \left( au\right)
\right) \right]  \eqnum{13}
\end{equation}

which consists of superposed impulse and shock waves. This latter term
arises due to the presence of the axion as can easily be seen from the Weyl
scalars given in Appendix A. We notice that there are no scalar curvature
singularities in this region which will make the Weyl scalar $\Psi _{4}$
divergent. However, $au=\frac{\pi }{2}$ is a coordinate singularity, or a CH
of type I [6] and as in the pure em problem it is a singularity of
quasiregular type [11,12].

Similarly, from the left (the $v$-dependent, Region III) we have the
corresponding expressions

\[
ds^{2}=2dudv-\cos ^{2}\left( bv\Theta (v)\right) \left[ dy^{2}+\left(
dx+\sin \left( bv\Theta (v)\right) dy\right) ^{2}\right] 
\]

\begin{equation}
A_{\mu }(v)=\frac{1}{\sqrt{2}}\sin \left( bv\Theta \left( v\right) \right) %
\left[ \delta _{\mu }^{x}+\delta _{\mu }^{y}\sin \left( bv\Theta (v)\right) %
\right]  \eqnum{14}
\end{equation}

\[
\kappa (v)=-\sin \left( bv\Theta (v)\right) 
\]

and the gravitational wave component

\begin{equation}
\Psi _{0}\left( v\right) =\frac{b}{2}\left[ i\delta \left( v\right) +b\Theta
(v)\left( \cos ^{2}\left( bv\right) -3i\sin \left( bv\right) \right) \right]
\eqnum{15}
\end{equation}

This region obviously shares the common features with those of region II.

\bigskip The foregoing Cauchy data on the intersecting null surfaces is
well-posed, therefore the two combinations of ``em+axion+gravity'' (with
trivial dilaton) collide at $u=0=v$, giving rise to the metric (2) and
fields (6) as the solution of the dynamical equations. The boundary
conditions to be imposed at the boundaries are those valid for the pure em
problem, namely the O'Brien-Synge conditions [14]. From the Weyl scalars
(Appendix A) we observe that only available singularities are the
distributional ones on the null boundaries. Namely,$\left( u=0,\text{ }bv=%
\frac{\pi }{2}\right) $ and $\left( v=0,\text{ }au=\frac{\pi }{2}\right) $
which \ occur also in the pure em problem. Off the boundaries $\left(
u>0,v>0\right) $ the spacetime is regular with the CH of type II [6], along $%
au+bv=\frac{\pi }{2}.$

\bigskip Finally we show the AdS structure (or BR form) of our metric as
follows. First we rewrite our metric in the form

\begin{equation}
ds^{2}=\frac{1}{2ab}\left( \frac{d\tau ^{2}}{\Delta }-\frac{d\sigma ^{2}}{%
\delta }\right) -\Delta dy^{2}-\delta \left( dx+\tau dy\right) ^{2} 
\eqnum{16}
\end{equation}

Next, we scale $x$ and $y$ by $\frac{1}{\sqrt{2ab}}$ and absorb the factor $%
2ab$ into $ds^{2}$. Now we apply the transformation

\[
\tau =\frac{1}{2r}\left( r^{2}-t^{2}+1\right) 
\]

\[
\sigma =\cos \theta 
\]

\begin{equation}
\tanh y=\frac{1}{2t}\left( r^{2}-t^{2}-1\right)  \eqnum{17}
\end{equation}

\[
x=\varphi -\frac{1}{2}\ln \left| \frac{\left( r+t\right) ^{2}-1}{\left(
r-t\right) ^{2}-1}\right| 
\]

and obtain

\begin{equation}
ds^{2}=\frac{1}{r^{2}}\left( dt^{2}-dr^{2}\right) -d\theta ^{2}-\sin
^{2}\theta \left( d\varphi -\frac{dt}{r}\right) ^{2}  \eqnum{18}
\end{equation}

which is in the required form of AdS. Similar extensions of the BR metric
were considered in Ref.s [15,16]. It has been shown in these references that
the near horizon geometry of an extreme Kerr black hole for large $r$ and
near the polar axis takes this form.

\section{Conclusion}

New solution to a system of Einstein-Maxwell-Dilaton-Axion, in the limit of
zero dilaton, is found. The problem is formulated as a problem of colliding
waves with physically well-defined Cauchy data. Interesting feature of the
solution is that it is singularity free. In the limit of vanishing axion it
reduces to the BS solution of colliding em waves. A transformation of our
metric casts it into a form that the AdS structure, which has deep
connection with conformal field theory, becomes manifest.

\section{Appendix A}

Null basis 1-forms of the Newman-Penrose formalism is chosen as

\[
l=du,\text{ \ \ \ }n=dv 
\]

\begin{equation}
\sqrt{2}m=\sqrt{\delta }dx+\left( i\sqrt{\Delta }+q_{0}\tau \sqrt{\delta }%
\right) dy  \eqnum{A.1}
\end{equation}

The Ricci components and the Weyl scalars are (the step functions in $%
a\Theta \left( u\right) $ and $b\Theta \left( v\right) $ are suppressed and $%
q_{0}$ is preserved. The Dirac delta functions are denoted by $\delta \left(
u\right) $ and $\delta \left( v\right) $ while other notations are as in Eq.
3).

\[
\Phi _{00}=b^{2}\left( 1-\frac{1}{4}\delta q_{0}^{2}\right) 
\]

\[
\Phi _{22}=a^{2}\left( 1-\frac{1}{4}\delta q_{0}^{2}\right) 
\]

\begin{equation}
\Phi _{02}=\Phi _{20}=ab\left( 1-\frac{1}{2}\delta q_{0}^{2}\right) 
\eqnum{A.2}
\end{equation}

\[
\Phi _{11}=-\frac{1}{8}ab\delta q_{0}^{2} 
\]

\[
\Lambda =\frac{1}{24}ab\delta q_{0}^{2} 
\]

\bigskip \bigskip

\bigskip 
\[
\Psi _{0}=-b\delta \left( v\right) \left[ \tan \left( au\right) -\frac{i}{2}%
q_{o}\cos \left( au\right) \right] +\frac{1}{2}q_{0}b^{2}\left( q_{0}\delta
+3i\sigma \right) 
\]
\begin{equation}
\Psi _{2}=\frac{1}{6}abq_{0}\left( q_{0}\delta +3i\sigma \right)  \eqnum{A.3}
\end{equation}

\begin{center}
\[
\Psi _{4}=-a\delta \left( u\right) \left[ \tan \left( bv\right) +\frac{i}{2}%
q_{0}\cos \left( bv\right) \right] +\frac{1}{2}q_{0}a^{2}\left( q_{0}\delta
+3i\sigma \right) 
\]
\end{center}

\section{Appendix B}

\begin{eqnarray*}
8\pi T_{\mu \upsilon }^{total} &=&\Phi _{00}n_{\mu }n_{\upsilon }+\Phi
_{22}l_{\mu }l_{\upsilon }+\Phi _{02}\left( m_{\mu }m_{\upsilon }+\overline{m%
}_{\mu }\overline{m}_{\upsilon }\right) +\left( \Phi _{11}+3\Lambda \right)
\left( l_{\mu }n_{\upsilon }+l_{\upsilon }n_{\mu }\right) \\
&&+\left( \Phi _{11}-3\Lambda \right) \left( m_{\mu }\overline{m}_{\upsilon
}+m_{\upsilon }\overline{m}_{\mu }\right)
\end{eqnarray*}

\begin{eqnarray}
&=&F_{\mu \alpha }F_{\text{ \ \ \ }\upsilon }^{\alpha }+\frac{1}{4}g_{\mu
\upsilon }F_{\alpha \beta }F^{\alpha \beta }+\frac{1}{4}\left( \kappa _{,\mu
}\kappa _{,\upsilon }-\frac{1}{2}g_{\mu \upsilon }\left( \nabla \kappa
\right) ^{2}\right)  \eqnum{B.1} \\
&=&4\pi T_{\mu \upsilon }^{em}+2\pi T_{\mu \upsilon }^{A}  \nonumber
\end{eqnarray}

The non-vanishing components of $T_{\mu \upsilon }^{em}$, $T_{\mu \upsilon
}^{A}$ and $G_{\mu \upsilon }$ are all listed below

\[
8\pi T_{uu}^{em}=a^{2}\left( 1+\sigma ^{2}\right) 
\]

\[
8\pi T_{vv}^{em}=b^{2}\left( 1+\sigma ^{2}\right) 
\]

\begin{equation}
8\pi T_{xx}^{em}=ab\delta \left( 1+\sigma ^{2}\right)  \eqnum{B.2}
\end{equation}

\[
8\pi T_{xy}^{em}=ab\tau \delta \left( 1+\sigma ^{2}\right) 
\]

\[
8\pi T_{yy}^{em}=ab\left( -1+2\tau ^{2}-\sigma ^{2}+\delta \tau ^{2}\sigma
^{2}\right) 
\]

\bigskip

\[
8\pi T_{uu}^{A}=a^{2}\delta 
\]

\[
8\pi T_{vv}^{A}=b^{2}\delta 
\]

\begin{equation}
8\pi T_{xx}^{A}=-ab\delta ^{2}  \eqnum{B.3}
\end{equation}

\[
8\pi T_{xy}^{A}=-ab\tau \delta ^{2} 
\]

\[
8\pi T_{yy}^{A}=-ab\delta \left( \Delta +\tau ^{2}\delta \right) 
\]

\bigskip

\[
G_{uu}=\frac{1}{4}a^{2}\left( \delta -4\right) 
\]

\[
G_{vv}=\frac{1}{4}b^{2}\left( \delta -4\right) 
\]

\begin{equation}
G_{xx}=\frac{1}{4}ab\delta \left( 3\delta -4\right)  \eqnum{B.4}
\end{equation}

\[
G_{xy}=\frac{1}{4}ab\tau \delta \left( 3\delta -4\right) 
\]

\[
G_{yy}=\frac{1}{4}ab\left[ \delta \left( 3\tau ^{2}\delta -4\right) +\Delta
\left( 3\delta +4\right) \right] 
\]

\end{document}